\documentstyle{article}

\begin{document}

{\bf A NEW APPROACH TO THE GEOMETRIZATION OF MATTER }

\bigskip

by: Louis Crane, Mathematics Department, KSU

\bigskip

{\bf ABSTRACT:} {\it We show that the sum over geometries in the
Lorentzian 4-D state sum model proposed for quantum GR in [1] includes
terms which correspond to geometries on manifolds with conical
singularities. Natural approximations suggest that they can be
interpreted as gauge bosons for the standard model, plus fermions,
plus dark matter.}

\bigskip

{\bf 1. Introduction}

\bigskip

Over the last few years, a new model for the quantum
theory of gravity has appeared. The model we are
referring to is 
in the class of spin foam models; more specifically it is a Lorentzian
categorical state sum model [1]. It is based on state sums on a
trangulated manifold, rather than differential equations on a smooth
manifold. The model has passed a number
of preliminary mathematical hurdles; it is actually finite on any
finite triangulation [2]. The biggest hurdle it still has to overcome is
an explicit physical interpretation, or differently put, a classical
limit.

The purpose of this paper is to outline a radically new way to
include matter in this type of theory. While it may seem premature
in light of the abovementioned hurdle, it is extremely natural in the
setting of the model, and perhaps easier to find than the classical
limit itself. If one accepts the approximate arguments we make, the
bosonic part of the standard model, rather than any
random collection of matter fields, is what appears. The approach
yields a fermionic sector as well, but we do not yet understand it.
Also a natural family of candidates for dark matter appears in it.
The crucial point of departure for this paper is the observation that
for a discrete state sum, unlike for a Lagrangian composed of
continuum fields, there is no need for the spacetime to consist
entirely of manifold points. We find that in investigating the kind of
singular points which the model naturally allows a number of
intriguing parallels to the standard model arise.

The realization that a specific possibility appears for including matter in
the model came as two complimentary points of view on the construction
of the model met. The first, the quantum geometric point of view, is
an interpretation of the categorical state sum as a sum over
Lorentzian discrete quantum geometries [1]. The second, the Group Field
Theory picture [3], interprets the state sums on particular
triangulations as Feynman diagrams for a quantum field theory on a
group manifold. The cross fertilization of these two approaches, as
discussed in [2] was an
important motivation for the finiteness proof. The sum over
Feynman diagrams in the GFT picture can be interpreted as a
superposition of quantum geometries in the LCSS point of view.

However, there is an important discrepancy between the two
pictures. Not every Feynman diagram in the GFT picture corresponds to
a manifold. The most general diagram, as we shall discuss below, is a
manifold with conical singularities over surfaces propagating along
paths and connecting at vertex-like cobordisms of the surfaces.

Thus we are faced with a dilemma. We must take one of the following
paths:

\bigskip

1.Restrict the class of Feynman diagrams we sum over in the GFT
  picture in a nonlocal and unnatural way,

\bigskip

2. Abandon the GFT picture altogether, and try to add matter to the
   LCSS picture; or

\bigskip

3. Reinterpret the conical singularities in the GFT picture as matter.

\bigskip

Thus, within the line of development we are pursuing, the new
proposal for matter, namely that it results from geometric excitations
at conical singularities, is actually the most parsimonious (as well, needless
to say, as the most optimistic) possibility to consider.

Put differently, the GFT picture seems to be telling us to take the
departure of the LCSS model, namely substituting a superposition of 
discrete geometries for a continuum picture, to its logical conclusion
of including all simplicial complexes.

The development of the LCSS approach to quantum gravity has proceeded,
rather surprizingly, at the mathematical level of rigor. The
finiteness result cited above is a theorem. Regrettably, the proposal in this
paper cannot be formulated as rigorously at this point. We are only
able to progress by making approximations. However,
it is at least possible to state the results as  conjectures, which a
careful study of some well defined integrals could in the future prove.

Since we are making a radical departure from existing lines of
development towards a fundamental understanding of matter fields, we
preface our proposal with a brief historical discussion, which shows
that the new suggestion is not as far conceptually from other
approaches as might at first appear. This is what motivates the phrase
``geometrization of matter'' in our title.

\bigskip

{\bf 2. Matter and Space}

\bigskip

Our current understanding of the physics of matter is rooted in the
idea of symmetry. Fields in quantum field theories are determined by
their quantum numbers, which index representations of the symmetries
of the theory. Interactions (vertices in Feynmanology) are linear maps
on state spaces which intertwine the action of the symmetries, or as
physicists like to say ``are not forbidden by the symmetries of the
theory.''

Our ideas about symmetry are much older than quantum field theory and
derive from our experience of space. Already in the nineteenth
century, mathematicians had the idea that different types of geometry
correspond to different types of symmetry.

When mathematicians and physicists have tried to understand the
symmetries of quantum physics, they have invariably resorted to
explaining matter fields in terms of one or another sort of
geometry. Aside from the manifestly spacetime symmetries of spin and
energy-monentum, every approach to find a fundamental explanation of
the internal or gauge symmetries has invoked one or another geometric
setting.

Thus gauge theory is formulated as the geometry of vector bundles,
Kaluza Klein theories resort to higher dimensional spacetime,
supergravity is based in superspace, string theory originally lived in
the geometry of loop spaces, while its M theoretic offspring seem to
be dwelling in bundles over manifolds of various dimensions again,
perhaps with specified submanifolds as well.

One could also mention noncommutative geometry, which studies
deformations both of families of symmetries and of the spaces they act
on to noncommutative C* algebras.

Nevertheless, at this point, we cannot say that any of these
approaches have really succeeded. A particular difficulty in many of
them has been the
failure of the standard model to emerge from a limitless set of
possibilities.

We want to propose that this historical survey suggests the following
points:

\bigskip

{\bf 1.Our understanding of symmetry is so rooted in geometry that if the
  fundamental theory of matter is not geometrical we will not find it
  anytime soon.}

\bigskip

 {\bf 2, We need to try a different type of geometry, and hope to get lucky
as regards the standard model.}

\bigskip

Now we want to claim that the ideas of quantum geometry which have
developed in the process of understanding the LCSS/GFT models point to
a natural generalization of the geometry of spacetime, namely
simplicial complexes. In this new geometric framework, it
is plausible that the standard model emerges naturally. To reiterate, the shift from
smooth manifolds to simplicial complexes is natural because we have substituted combinatorial state sums for differential
equations.

\bigskip

{\bf 3. The topology of a class of simplicial complexes.}

\bigskip

The GFT picture is a generalization of the idea of a state sum
attached to a triangulation of a 4-manifold. The picture
is to think of a triangulation of a 4-manifold as a 5-valent graph
with each edge of the graph refined into a bundle of 4
strands. Different matchings of the strands at an edge are different
diagrams.
The
vertices of the diagram correspond to the 4-simplices, the edges are the 3-simplices, and the
strands are the faces of the 3-simplices. Mathematicians would
describe 
this as the dual 2-skeleton of the triangulation, following strands
around until they close into loops, and attaching disks along the loops.

A crucial observation is that the LCSS model in [1], unlike the
topological models that preceeded it [4], requires only the combinatorial
data of the dual 2-skeleton to formulate it, since it has no terms on
the edges or vertices of the triangulated manifold. The GFT picture
actually goes beyond this and produces all possible strand diagrams
as terms in an expansion of a field theory into Feynman graphs.

A standard argument from PL topology tells us when the complex we
would build up from such a diagram by adding simplices of dimensions 1
and 0 is a manifold: the links of all simplices must be spheres of
appropriate dimension. (The combinatorial picture described above in
fact tells us how they should be added). In the situation we are considering, this will
be automatically satisfied for the 4- 3- and 2- simplices, the link of
a 1-simplex can be any 2-manifold, and the link of a vertex can be any
3-manifold with conical singularities on the surfaces corresponding to
the links of the incident edges. The links of vertices can also be
described as 3-manifolds with boundary components the links of the
edges incident to the vertex, leaving the cones out for simplicity.

For the nonmathematical reader, we note that a cone over any space is
the cross product of an interval with the space, with the copy at one
end of the interval contracted to a point. A point in a manifold has
neighborhoods which are homeomorphic to the cone over a sphere, which
is just a ball. A
point with a neighborhood homeomorphic to a cone over some other
manifold is not a manifold point, and is referred to as a conic
singularity. In the case of an isolated singularity, the submanifold
over which the cone is constructed is called the link of the point. If
instead we have a simplex crossed with a cone on a lower dimensional
submanifold, it is the link of the simplex. In a triangulated
manifold, all links of simplices are spheres of appropriate dimension.

The proposal we are making is to interpret the web of singularities
in such a complex as a Feynman graph; that is to say, we want to
interpret the low energy part of the geometry around the cones over
surfaces as particles, and the 3-manifolds with boundary connecting them as
interaction vertices.

We shall make an attempt at this below, using several approximate
techniques. At this point, we wish to underscore the extreme
parsimoniousness of this proposal. Nothing is added to the model for
quantum GR, no extra dimensions, no larger group, We simply allow a
natural larger set of configurations. Once we abandon smooth manifolds
for PL ones, there is really no reason not to allow such configurations. 
In the GFT picture, where geometries appear as fluctuations of a
nongeometric vacuum, they are on an equal footing.

\bigskip

{\bf 4. Conical matter}

\bigskip

Now we want to get some picture of what degrees of freedom would appear
on a conical singularity in the model of [1]. Since we need to average
over all triangulations this is not easy. Also, in order to obtain a
model to compare with particle physics as we see it today, we need to
describe a universe which has cooled enormously from the Planck
temperature, i.e. we need a low energy limit of the model. At present
we do not know how to abstract such a limit from the model directly,
so we approach this problem by
making use of the connection between the discrete models for TQFTs  in [4] and
the model in [1].

A TQFT is an automatic solution to the renormalization group, in the
sense that if we make a refinement of the triangulation on which it is
computed the result is unchanged. We want to suggest
that the state space of low energy states which survive summing over refinements
of triangulation of a cone over a surface is given by the space of
states for an associated 2+1 dimensional TQFT on the surface with a
puncture. (The puncture would allow information to flow out, thus
imitating the conical singular point. The state space for a TQFT with
puncture is larger than the one for a closed surface in a TQFT.)

This should be taken as a physical hypothesis at this moment. One
reason for believing it is that the LCSS model in [1] is itself
obtained by constraining a TQFT.

Then there is the question of what TQFT to expect. Since the state sum
in [1] is from the unitary representations of SL(2,C), which is a sort
of double for Su(2), the TQFT should be the one for $ SU_q(2) \times
SU_q(2)$, i.e. a left-right symmetric TQFT produced from a quantum
group in the by now standard way. This is also the TQFT we constrain
to produce the euclidean signature model for quantum general
relativity in [5].

Approximating limits of theories by states of other theories is not an
unknown technique in theoretical physics. At this point we do not know
how to set q or what value it should take. We could introduce a q into
our original model by passing to the 
Quantum Lorentz Algebra [7]. It
may also be that a q emerges from the poorly understood limit of low
energy in the model, as a cosmological constant. As we note below,
certain choices for q have interesting implications for the particle
content of the low energy theory.

We believe that in the future it may be possible to make a stronger
argument for this. The reason has to do with the relationship between  
conformal structures on a surface and flat Lorentzian metrics on the
cone over the surface. A Riemann surface can be obtained by
quotienting the hyperboloid in three dimensional Minkowski space by a
discrete subgroup of the 2+1 dimensional Lorentz group, which is
isomorphic to SL(2,R). Quotienting the entire forward timelike cone by
the same group yields a flat Lorentzian metric on the entire cone over
the surface, except that the conical singularity (the origin in
Minkowski space), is not a manifold point, so naive definitions of
metrics fail there. Thus, the approach to producing CSW theory by
quantizing a bundle over Riemann moduli space could be interpreted as
a quantization over the space of flat geometries around a conic
singularity. States arising from effects around flat geometry should
be important in understanding the low energy behavior of the model. 
This argument will be difficult because it will be
necessary to treat the effect of the singular point, so we do not
attempt it here. We will make further use of the relationship between flat
Lorentzian metrics on a cone and constant negative sectional curvature
(hyperbolic) metrics on the
boundary of the cone in what follows.

At this point, let us note that the space of states assigned to a once
punctured torus by a TQFT is a very special object. As demonstrated in
[6], it is always a Hopf algebra object in the category associated to
the TQFT. In the case of the TQFT produced from $SU(2)_q$, also known
as the CSW [8] model, it is a sum of matrix rings, one at each
dimension, up to the cutoff determined by q. The unitary part of this
has been suggested by Connes and Lott [9] as a natural origin for the
gauge symmetry of the standard model.

Thus, according to our ansatz using TQFT states, we find a
copy of the gauge bosons of the standard model in the states on a
toroidal conical singularity. If we choose the q in such a way as to
get exactly 3 matrix blocks in our space [10], we could get exactly the
standard model, otherwise we could be led to the conjecture that the
standard model is really part of a gauge theory with group U(1)+SU(2)+
SU(3)+SU(4)... where particles charged in the higher dimensional
pieces acquire very large masses and are therefore unseen.

It is therefore interesting to ask what sort of interaction
vertices toroidal and other conical singularities might admit. Are the
toroidal singularities special, as compared to the higher genus ones?

\bigskip

{\bf 5. Hyperbolic manifolds and interaction vertices for conic
matter.}

\bigskip

We remind the reader that we are interpreting regions which look like
a conic singularity over a surface crossed with an interval as
propagating particles. Now we want to think of the vertices where such
topologies meet as interaction vertices. As we explained above, the
regions around these vertices are cones over 3-manifolds with conic
singularities over surfaces. 

We now want to propose a second approximation. The low energy vertices
corresponding to these cones over 3-manifolds should be dominated by
the flat Lorentzian metrics on them. The physical argument justifying
this is that topologies which did not admit flat geometries would
become very high energy as we summed over refinements of the
triangulation. In the related context of 3d manifolds discussed above
we noted a possible connection between this approximation and the TQFT
ansatz.

Now we discover an interesting connection. Flat Lorentzian geometries
on the cone over a 3-manifold arise in a natural way from hyperbolic
structures on it. This is because hyperbolic structures can be
recovered as the quotient space of the forward timelike hyperboloid of
Minkowski space by discrete subgroups of SL(2,C) acting isometrically
on it. Extending the action to the entire forward cone yields a flat
Lorentzian 4-geometry on the cone. If we do a similar construction to
produce a 3-manifold with boundary, we obtain a conformal
(=hyperbolic) structure on 
the two dimensional boundary of the 3-manifold at the same time.
Thus we are led to a picture where we match the hyperbolic structures
on the surfaces linking the edges to the hyperbolic structures
assigned to the boundary components of the 3-manifolds linking the
vertices to obtain flat geometries surrounding the entire singular
part of a 4-D simplicial complex which could arise in our model.

An interesting theorem about hyperbolic structures on 3-manifolds with
boundary, called Mostow rigidity [11], tells us that the degrees of freedom
of a hyperbolic structure on the bulk are exactly the degrees of
freedom of the conformal structure on the boundary components. This
means that when we sum over flat geometries in our situation, we get a
multiple integral over Teichmuller parameters. This produces a sort of
mathematical convergence with the Polyakov approach to string
theory. We do not yet know if when we go to quantizing over the space
of flat structures  any deeper connections to string theory will
result.

Now we make another critical observation: the only complete hyperbolic
3-manifolds with finite volume  are the ones whose boundary components
are tori and Klein bottles [12]. We believe that infinite volume metrics
would not make an important low energy contribution to the model,
while incomplete hyperbolic metrics would not match flatly at the
surfaces linking the edges.

This leads us to a picture in which the low energy interacting world
would contain only toroidal and Klein bottle singularities, leaving
the higher genus surfaces to decouple and form dark matter. Since our
TQFT ansatz suggested that the states on tori could reproduce the
gauge bosons for the standard model, while the Klein bottle, being
nonorientable, would produce fermionic states, this yields a 
picture with many similarities to the standard model plus dark
matter. We have not yet tried to find an argument for the state space 
on a Klein bottle. 

We would also like to find an approximate argument for how TQFT states
might propagate across a vertex described by some cobordism between
the incoming and outgoing surfaces. The most obvious would be to
simply take the linear map between the surface states given by the
TQFT itself. It is interesting to note that for a particularly simple
cobordism from two tori to a third this would just give the
multiplication of the associative algebra we mentioned above, yielding
the gauge algebra of the standard model.

\bigskip

{\bf Conclusions}

\bigskip

It is clear that the arguments presented here to analyse the behavior
of the LCSS model near singular points are very preliminary; it would
be rash to jump to the conclusion that we had the unified field theory
in hand. Nevertheless the trilogy of standard model bosons, fermions,
and weakly interacting higher genus states, especially embedded in a
plausible model for quantum general relativity, cannot be ignored.

It does seem safe to say that the simplicity of this model makes it an
interesting problem for mathematical physics to study. The connection
between particle interactions and hyperbolic structures on 3-manifolds
has the advantage that it poses problems for a mathematical subject
which has been deeply studied from several points of view and
concerning which much is known [12]. It is interesting to note that
the question of hyperbolic structure on 3-manifolds with toroidal
boundaries is deeply connected to knot theory. We may find knotted
vertex structures play a role in this theory; interestingly, they are
chiral.

Given the finiteness proof in [2], the conjectures in this paper
pertain to limits of families of finite integrals; at least in
principle there is reason to hope they can be rigorously formulated
and proven.

The fact that the families of flat
structures which appear here are parametrized by the Teichmuller
parameters on the boundary (Mostow rigidity) means that the subject
takes on an unexpected mathematical resemblance to string theory,
although it is ``world sheets'' rather than loops which propagate
through spacetime.

We do think the moral can be drawn from this model that there are more
possibilities for forming fundamental quantum theories of nature than
contemporary theoretical physics seems to recognize.

\bigskip

BIBLIOGRAPHY

\bigskip

[1] J. Barrett and L. Crane, A Lorrentzian Signature Model for Quantum
General Relativity CQG 17 (2000) 3101-3118.

\bigskip
see also:

\bigskip

J. C. Baez, Spinfoam Models CQG 15 (1998) 1827-1858.

\bigskip

[2]A. Perez and C. Rovelli A Spin Foam Model without Bullbe
Divergences Nucl hys B 599 (2001) 427-434

\bigskip

[3] L. Crane  A. Perez, C. Rovelli, A Finiteness Proof for the
Lorentzian State Sum Model for General Relativity gr-qc 0104057

\bigskip

[4] L. Crane L. Kauffman and D. Yetter, State Sum invariants of
4-Manifolds JKTR  (2) (1997)  177-234.

\bigskip

[5] J. Barrett and L. Crane Relativistic Spin Nets and Quantum
Gravity, J. Math. Phys 39 (1998) 3296-3302.

\bigskip

[6] L. Crane and D. Yetter, On Algebraic Structures Implicit in
Topological Quantum Field Theories, JKTR 8 (2) (1999) 125-164.

\bigskip

[7] E. Buffenoir and P. Roche, Harmonic Ananysis on the Quantum Group,
CMP 207, (1999) 499-555.

\bigskip

[8] E. Witten, Quantum Field Theory and the Jones Polynomial, CMP 121
(1989) 351-399.

\bigskip

[9] A. Connes and J. Lott Particle Models and Non-Commutative Geometry
Nucl. Phys. B (Proc. Supp.) 18B 29 (1990)

\bigskip

[10] R. Coquereaux, A. O. Garcia and R. Trinchero, Differential
Calculus and Connections on a Quantum Plane at a Cubic Root of Unity,
 Rev. Mod. Phys. 12 (2000) 227-285

\bigskip

[11] G. D. Mostow, Strong Rigidity os locally Symetric Spaces, Ann
Math. studies no. 78 PUP 1973

\bigskip

[12] W. Thurston The Geometry and Topology of Three-Manifolds
Princeton University notes, unpublished

\end{document}